\documentclass[reprint,twocolumn,showpacs,amssymb,amsmath,aps,pra]{revtex4-1}
\newcommand{\half}{\mbox{$\frac{1}{2}$}}
\usepackage{bm}
\usepackage{graphics}
\usepackage{amsmath}
\usepackage{amssymb}
\usepackage{pstricks}
\usepackage{multido}
\usepackage{epsfig}
\begin{document}
\title{Quantum discord in finite XY chains}
\author{L. Ciliberti, R. Rossignoli, N. Canosa}
\affiliation{Departamento de F\'{\i}sica-IFLP,
 Universidad Nacional de La Plata, C.C. 67, La Plata (1900) Argentina}
\date{\today}
\begin{abstract}
We examine the quantum discord between two spins in the exact ground state of
finite spin $1/2$ arrays with anisotropic $XY$ couplings in a transverse field
$B$. It is shown that in the vicinity of the factorizing field $B_s$, the
discord approaches a common finite non-negligible limit which is independent of
the pair separation and the coupling range. An analytic expression of this
limit is provided. The discord of a mixture of aligned pairs in two different
directions, crucial for the previous results, is analyzed in detail, including
the evaluation of coherence effects, relevant in small samples and responsible
for a parity splitting at $B_s$. Exact results for finite chains with first
neighbor and full range couplings and their interpretation in terms of such
mixtures are provided.
\end{abstract}
\pacs{03.65.Ud,03.67.Mn,75.10.Jm}
\maketitle
\section{Introduction}
The great interest on quantum entanglement in recent years was triggered by the
key role it played in certain quantum processing tasks like quantum
teleportation and superdense coding \cite{QT.93,BW.92,NC.00}. It was also shown
to be essential for achieving an exponential speedup over classical computation
in pure-state based quantum computation \cite{JL.03}. However, it was recently
recognized that in the case of mixed-state based quantum computation, like the
``deterministic quantum computation with one qubit'' model introduced by Knill
and Laflamme \cite{KL.98}, an exponential speed up can take place without a
substantial presence of entanglement \cite{DFC.05}. This has turned the
attention to alternative measures of quantum correlations in mixed states, like
the quantum discord, introduced by Ollivier and Zurek \cite{Zu.01}, which can
detect those quantum correlations present in certain separable mixed states
\cite{RW.89} and hence not captured by the entanglement of formation
\cite{Sch2}, but still useful and crucial for certain quantum tasks. It was in
fact recently shown by Datta, Shaji and Caves \cite{Ca.08} that the circuit of
\cite{KL.98} does exhibit a finite non-negligible value of the quantum discord
between the control qubit and the remaining mixed qubits. Since then, interest
on the quantum discord and other alternative measures has grown considerably
\cite{Luo.08,Dat.09,SL.09,ARA.10,Ve.10} and several studies of their behavior
in spin pairs immersed in a spin chain have been performed
\cite{Dd.08,Sar.09,WR.10,MG.10}.

The aim of this work is to analyze the quantum discord of spin pairs in the
exact ground state of {\it finite} spin $1/2$ arrays with anisotropic $XY$ or
$XYZ$ type couplings in a uniform transverse magnetic field $B$. The exact ground state of a finite chain will have 
a definite spin parity and this property will be seen to deeply affect the
discord for fields lower than the critical field $B_c$, where we will show that the main results can be interpreted in terms of the discord of mixtures of aligned pairs. Our interest in these systems is motivated in particular by the remarkable factorization  phenomenon they can exhibit at 
a particular finite value $B_s<B_c$ of the magnetic field  \cite{Kur.82,Ros.04,Ros.05,Am.06, 
RCM.08,GAI.09,RCM.09,GAI.10}: At such field, they have an exactly {\it separable} (i.e., factorized) ground state. 
This feature was first 
discovered in \cite{Kur.82} in one-dimensional (1-$d$) $XY$ chains with first neighbor couplings, and later shown to occur also in more general systems, like 2-$d$ arrays \cite{Ros.05}, cyclic \cite{RCM.08} and general \cite{GAI.09,RCM.09} arrays with arbitrary range couplings with a common anisotropy and also systems in non-uniform fields \cite{RCM.09}. 
For transverse fields, these separable ground states actually break parity symmetry and are hence degenerate, coinciding $B_s$ in a finite system with the last crossing of the two lowest opposite parity levels \cite{RCM.08}.  
A most remarkable related feature is that in the immediate vicinity of $B_s$, pairwise entanglement, though weak, reaches {\it full range} \cite{Am.06,RCM.08}, regardless of the coupling range \cite{RCM.08}, 
changing at $B_s$ from antiparallel ($B<B_s$) to parallel $(B>B_s)$ type \cite{Am.06}, indicating an ``entanglement transition''. This suggests the possibility of a non-zero discord between distant pairs at least in the vicinity of $B_s$, with universal features (independence of separation and coupling range). Here we will show that this is indeed the case, and derive also its analytic limits at this field. 

Moreover, distributed pairwise 
entanglement is necessarily weak due to the well-known monogamy property \cite{CKW.00,OV.06} associated with entanglement sharing: If in a system of $n$ spins or qubits one spin is strongly entangled with a second spin, it cannot be strongly entangled with any of the remaining spins. This fundamental feature follows from the bound \cite{CKW.00,OV.06} $\sum_{j}C_{ij}^2\leq C_i^2\leq 1$ satisfied by the pairwise concurrences \cite{Wo.98} $C_{ij}$ measuring the entanglement between spins $i$ and $j$, where $C_i$ represents the concurrence between $i$ and the rest of the chain. If all pairs are equally entangled, the maximum value that can be reached by $C_{ij}$ is in fact just $2/n$ \cite{KBI.00,Du.01}.    
In contrast, the quantum discord is not affected by such bound and can simultaneously reach non-negligible finite values between any two spins, 
as will be seen to occur in the vicinity of $B_s$, leading to a quite different behavior with the applied field and separation in the whole region $|B|<B_c$. The properties of the ground state discord in the vicinity of the separability field in anisotropic XY chains were not discussed in previous references. We will also analyze here finite size effects, which lead to a finite step of the discord at the factorizing field $B_s$ and other parity transitions, visible in small chains. 

Sec.\ \ref{II} discusses the quantum discord and its evaluation in typical
reduced states of a spin pair in such chains, describing in detail the case of
a mixture of two aligned states, which represents the exact reduced state in the vicinity of 
$B_s$. Coherence effects in these mixtures, relevant for small chains, are also
examined. Sec.\ \ref{III} discusses the behavior of the discord of spin pairs
with the applied field and separation in finite cyclic $XY$ chains with first
neighbor as well as full range couplings, including its interpretation 
and the differences with the pair entanglement.
The appendix discusses the details of the exact definite parity solution of the
finite cyclic $XY$ chain. Conclusions are finally drawn in \ref{IV}.

\section{Formalism\label{II}}
\subsection{Quantum Discord}
The quantum discord ($D$) is a measure of quantum correlations based on the
difference between two distinct quantum generalizations of the classical mutual
information, or equivalently, the classical conditional entropy
\cite{Zu.01,Ca.08}. Given a bipartite system $A+B$ in a mixed state
$\rho_{AB}$, and denoting with $\rho_{A}={\rm Tr}_{B}\,\rho_{AB}$, $\rho_B={\rm
Tr}_A\,\rho_{AB}$ the reduced density operators of each subsystem, $D$ can be
expressed as \cite{Zu.01}
\begin{subequations}
\label{D}
\begin{eqnarray}
D&=&I(A:B)-\mathop{\rm Max}_{\{P^B_j\}}\,I_{\{P^B_j\}}(A:B)\label{D1}\\
&=&\mathop{\rm Min}_{\{P^B_j\}}[S(\rho'_{AB})-S(\rho'_B)]-
[S(\rho_{AB})-S(\rho_B)]\,,\label{D2}
\end{eqnarray}
\end{subequations}
where $I(A:B)=S(\rho_A)+S(\rho_B)-S(\rho_{AB})\geq 0$ is the quantum mutual
information, with $S(\rho)=-{\rm Tr}\,\rho\log_2\rho$ the von Neumann entropy,
and $I_{\{P^B_j\}}(A:B)=S(\rho_A)+S(\rho'_B)-S(\rho'_{AB})$ the mutual
information after a local von Neumann measurement in system $B$ defined by a
set of orthogonal projectors  $P_j^B=I_A\otimes |j_B\rangle\langle j_B|$. Here
\begin{equation}
\rho'_{AB}=\sum_j P^B_j\rho_{AB} P^B_j\label{rhopab}
\end{equation}
represents the joint state after such measurement if the result is unknown,
with $\rho'_B={\rm Tr}_A\,\rho'_{AB}$ and $\rho'_A={\rm Tr}_B\,\rho'_{AB}=\rho_A$.
 Minimization in (\ref{D2}) is over all
sets of local projectors.

The last bracket in (\ref{D2}) is the direct quantum extension of the classical
conditional entropy \cite{Wh.78,NC.00} ($S(A|B)=S(A,B)-S(B)=\sum_{j}p_j
S(A/j)$, with $S(A,B)=-\sum_{i,j}p_{ij}\log_2 p_{ij}$, $S(B)=-\sum_j
p_j^B\log_2 p_j^B$ and $S(A/j)=-\sum_j (p_{ij}/p_j^B)\log_2(p_{ij}/p_j^B)$ for
a system with joint probability distribution $p_{ij}$ and marginal distribution
$p_j^B=\sum_i p_{ij}$), which, in contrast with the classical case, is not
necessarily positive. On the other hand, the first bracket is the conditional
entropy $S_{\{P_j^B\}}(A|B)=\sum_j p_j S(\rho_{j})$ after a local measurement
in system $B$, where $p_j={\rm Tr}\,P^B_j\rho_{AB}$ is the probability of
outcome $j$ and $\rho_{j}= P_j^B\rho_{AB}P_j^B/p_j$ the state after such
outcome. It represents the average lack of information about $A$ after such
measurement and is a non-negative quantity. The second term in (\ref{D1}) is
considered a measure of the {\it classical} correlations between $A$ and $B$ in
\cite{HV.01}, the discord measuring then the quantum part of these
correlations.

In the case of a pure state ($\rho_{AB}^2=\rho_{AB}$), $S(\rho_{AB})=0$ while
the first bracket in (\ref{D2}) vanishes for any choice of local projectors,
and the discord reduces then to the entanglement entropy \cite{Sch1.95}
$D=E=S(\rho_A)= S(\rho_B)$. In the case of mixed states, however, it does not
coincide in general with the entanglement of formation \cite{Sch2} (the convex
roof extension of the entanglement entropy). While the latter vanishes for any
separable state, i.e., for a convex superposition of product states
$\rho_{AB}=\sum_{\alpha} q_\alpha\rho_A^\alpha\otimes \rho_B^\alpha$,
$q_\alpha\geq 0$ \cite{RW.89}, the discord can be non-zero for these states.
The discord vanishes in the case of densities diagonal in an orthogonal product
basis ($b_{AB}=\{|i_A\rangle |j_B\rangle\}$) or in general a conditional
product basis ($b_{AB}=\{|i_{jA}\rangle|j_B\rangle\}$, with the orthogonal set
$\{|i_{jA}\rangle\}$ depending on $|j_B\rangle$), but will not vanish in
general for a mixture of non-commuting product states, which is still a
separable state. The difference (\ref{D}) can be shown to be non-negative
\cite{Zu.01} due to the subtle concavity property of the {\it conditional} von
Neumann entropy $S(\rho_{AB})-S(\rho_B)$ \cite{Wh.78}.

\subsection{Quantum discord and entanglement of spin pairs in definite parity
 states}
Let us now describe the basic elements to evaluate the discord and entanglement
of a spin pair in a typical eigenstate of a finite chain of $n$ spins, where
the rest of the spins will play the role of an environment. We will consider
spin $1/2$ chains with $XYZ$ couplings of arbitrary range in a uniform
transverse magnetic field $B$ along the $z$ axis, such that the chain
Hamiltonian has the form
 \begin{equation}
H=B \sum_{i}s_{iz}-\half\sum_{i\neq j}\sum_{\mu=x,y,z}J^{ij}_\mu
s_{i\mu}s_{j\mu}\,,
 \label{H}\end{equation}
where $s_{i\mu}$ denotes the spin components at site $i$ (in units of $\hbar$).
$H$ commutes with the spin parity operator
\begin{equation}
P_z=\exp[i\pi\sum_i (s_{iz}+1/2)]=\prod_i(-\sigma_{iz})\,, \label{Pz}
\end{equation}
where $\sigma_{iz}=2s_{iz}$, which changes $s_{i\mu}$ to $-s_{i\mu}$ $\forall$
$i$ and $\mu=x,y$. Non-degenerate eigenstates $|\Psi_\nu\rangle$ of $H$ will
then have a definite spin parity $P_z=\pm 1$, a symmetry which will play a
fundamental role in our discussion.

The reduced density matrix of an arbitrary pair $i,j$ in such eigenstate,
\begin{equation}
\rho_{ij}={\rm Tr}_{n-ij}|\Psi_\nu\rangle\langle\Psi_\nu|\,,\label{rhoij0}
\end{equation}
will then contain no elements connecting states of opposite parity, commuting
therefore with the pair parity $P_z^{ij}=\sigma_{iz}\sigma_{jz}$. In the
standard basis $\{|00\rangle,|01\rangle,|10\rangle,|11\rangle\}$
($|kl\rangle\equiv|k_i\rangle|l_j\rangle$, with $s_{iz}|k_i\rangle=\half
e^{i\pi k_i}|k_i\rangle$, $k_i=0,1$), $\rho_{ij}$ will then be of the form
\begin{equation}
\rho_{ij}=\left(\begin{array}{cccc}a&0&0&\alpha\\0&c&\beta&0\\
 0&\bar{\beta}&c'&0\\\bar{\alpha}&0&0&b\end{array}\right)\,,
 \label{rij}\end{equation}
where, setting  $\langle O\rangle\equiv {\rm tr}\,\rho_{ij}\,O$
and $s_{i\pm}=s_x\pm is_{iy}$,
 \begin{eqnarray}
(^a_b)&=&{\textstyle\frac{1}{4}}
\pm\half\langle s_{iz}+s_{jz}\rangle+\langle s_{iz}s_{jz}\rangle',,
\label{vm1}\\
(^c_{c'})&=&{\textstyle\frac{1}{4}}\pm\half\langle s_{iz}-s_{jz}\rangle
-\langle s_{iz}s_{jz}\rangle\,,\label{vm2}\\
 (^\alpha_\beta)&=&\langle s_{i-}s_{j\mp}\rangle\,,
 \label{vm3}\end{eqnarray}
with $a+b+c+c'=1$. We will here consider translationally invariant systems such
that $\langle s_{iz}\rangle=\langle s_{jz}\rangle$ and hence $c=c'=
\half(1-a-b)$. Moreover, $\alpha$ and $\beta$ will be real since $H$ can be
represented by a real matrix in the standard product basis of the full space.
Non-negativity of $\rho_{ij}$ implies $|\alpha|\leq \sqrt{ab}$, $|\beta|\leq
c$, with $a,b,c$ non-negative.

The internal entanglement of the pair can be measured through the entanglement
of formation, which for the case of two qubits can be explicitly calculated as
\cite{Wo.98}
 \begin{equation}
 E=-\sum_{\nu=\pm}q_\nu\log_2 q_\nu
 \,,\;q_{\pm}=\half(1\pm\sqrt{1-C^2})\,,\label{Eij}
 \end{equation}
where $C$ is the {\it concurrence} \cite{Wo.98}. It is given here by
\begin{equation}
C=2\,{\rm Max}[|\alpha|-c,|\beta|-\sqrt{ab},0]\,.
 \label{Cij}\end{equation}
The entanglement of the pair is of parallel (antiparallel) type if the first
(second) entry is positive \cite{Am.06,RCM.08}. Just one of these entries can be
positive for a non-negative $\rho_{ij}$.

On the other hand, in order to evaluate the discord of the pair, we need first
the eigenvalues of $\rho_{ij}$, given for $c=c'$ by $\lambda_{ij}=
(\frac{1-a-b}{2}\pm |\beta|,\frac{a+b}{2}\pm\sqrt{
(\frac{a-b}{2})^2+|\alpha|^2})$, and those of the single spin density matrix,
 \begin{eqnarray}\rho_j&=&{\rm Tr}_{i}\rho_{ij}=
 \left(\begin{array}{cc}a+c&0\\0&c+b\end{array}\right)\,,
\end{eqnarray}
which are obviously $\lambda_j=\half[1\pm(a-b)]$. We also need to consider a
measurement of the spin at site $j$ along an arbitrary axis $z'$ defined by the
angles $\gamma$ and $\phi$. The state of the pair after such measurement (Eq.\
(\ref{rhopab})) is
\begin{equation}
\rho_{ij}'=P^j_{0'}\rho_{ij}P^j_{0'}+P^j_{1'}\rho_{ij}P^j_{1'}\,,
\label{rdij}
\end{equation}
where $P^j_{k'}=I_i\otimes|k'\rangle\langle k'|$ for $k=0,1$ with
\begin{eqnarray}|0'\rangle&=&
\cos(\half \gamma)|0\rangle+e^{i\phi}\sin(\half\gamma)|1\rangle\\
|1'\rangle&=& \cos(\half
\gamma)|1\rangle-e^{-i\phi}\sin(\half\gamma)|0\rangle\,,
 \label{rdija}\end{eqnarray}
such that $s_{jz'}|k'\rangle=\half e^{i\pi k}|k'\rangle$ for
$s_{jz'}=s_{jz}\cos\gamma+ s_{jx}\sin\gamma\cos\phi+s_{jy}\sin\gamma\sin\phi$.
For real $\alpha, \beta$, the eigenvalues of $\rho'_{ij}$ are
\begin{equation}
\lambda'_{ij}={\textstyle\frac{1+\nu(a-b)\cos\gamma+\mu
\sqrt{[(2(a+b)-1)\cos\gamma+\nu(a-b)]^2+
4|\bm{\alpha}+\bm{\beta}|^2\sin^2\gamma}}{4}} \label{lmn}
 \end{equation}
where $\nu=\pm 1$, $\mu=\pm 1$ and $|\bm{\alpha}+\bm{\beta}|^2 \equiv
\alpha^2+\beta^2+2\alpha\beta\cos 2\phi$, corresponding $\nu=1$ $(-1)$ to the
eigenvalues of the first (second) term in (\ref{rdij}). The eigenvalues of
$\rho'_j={\rm Tr}_{i}\,\rho'_{ij}$ are then
$\lambda'_j=\half[1\pm(a-b)\cos\gamma]$.

We have then all elements to evaluate the difference
\begin{equation}
D(\gamma,\phi)=S(\rho_{ij}')-S(\rho_j')-[S(\rho_{ij})-S(\rho_j)]\,,\label{DS}
\end{equation}
whose minimum (with respect to $\theta,\phi$) is the discord $D$ (Eq.\
(\ref{D})). For $\alpha\beta\geq 0$, minimization with respect to $\phi$ yields
$\cos 2\phi=1$ and just the minimization over $\gamma$ is finally required,
which can be restricted to the interval $[0,\pi/2]$. The minimum for the pair
densities used in sec.\ \ref{III} was obtained for $\gamma=\pi/2$ ($z'=x$),
where $\lambda'_{ij}=\frac{1+\mu\sqrt{(a-b)^2+ 4(\alpha+\beta)^2}}{4}$ becomes
independent of $\nu$ and hence degenerate. A general evaluation of the discord
for states of the form (\ref{rij}) was recently provided \cite{ARA.10}.

\subsection{The case of a mixture of two aligned states}
A particular case of (\ref{rij}) of exceptional interest is that of a
statistical mixture of two aligned states along arbitrary directions, not
necessarily opposite, such that the local states involved are {\it
non-orthogonal}. Choosing the $z$ axis as the bisector of the angle
between the two directions, we can write this state as
\begin{eqnarray}
\rho_{ij}(\theta)&=&\half(|\theta\theta\rangle\langle\theta\theta|+
|\!-\!\theta\!-\!\theta\rangle\langle\!-\!\theta\!-\!\theta|)\label{rij2a}\\
&=&\left(\begin{array}{cccc} a&0&0&\alpha\\0&\alpha&\alpha&0\\
0&\alpha&\alpha&0\\\alpha&0&0&b\end{array}\right)\,,\;
\begin{array}{l}(^a_b)=\frac{1}{4}(1\pm\cos\theta)^2\\
\alpha=\frac{1}{4}\sin^2\theta\end{array}\,,
 \label{rij2}\end{eqnarray}
where $|\theta\rangle=\exp[i\theta s_{iy}]|0\rangle= \cos\half\theta
|0\rangle+\sin\half\theta|1\rangle$ and (\ref{rij2}) is again the standard
basis representation. Eq.\ (\ref{rij2a}) is the exact reduced state of any two spins in the immediate vicinity of the factorizing field (see sec.\ \ref{III}) if coherence terms are neglected. It also provides the basic approximate picture of the pair state in the region $|B|<B_c$.

The state (\ref{rij2a}) is obviously {\it separable} \cite{RW.89}, i.e., a
convex combination of product states, and therefore its concurrence and
entanglement are identically 0, as verified from Eq.\ (\ref{Cij}). However, its
discord {\it is positive for $\theta\in(0,\pi/2)$}, vanishing just for
$\theta=0$ or $\theta=\pi/2$: If $\theta=0$, Eq.\ (\ref{rij2a}) becomes a pure
product state (and hence $D(\gamma,\phi)=0$ $\forall$ $\gamma,\phi$) whereas if
$\theta=\pi/2$, $|\theta\rangle$ and $|-\theta\rangle$ are {\it orthogonal} and
$\rho(\theta)$ becomes {\it diagonal in a product basis}, with Eq.\ (\ref{DS})
vanishing then for $\gamma=\pi/2$ and $\phi=0$ ($\rho'_{ij}=\rho_{ij}$).

\begin{figure}[t]

\centerline{\hspace*{0.25cm}\scalebox{.7}{\includegraphics{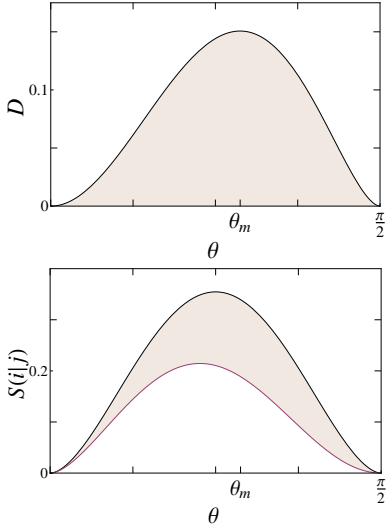}}}
\vspace*{-.25cm}

\caption{(Color online) Top: The quantum discord $D\equiv D(\pi/2,0)$ as a
function of $\theta$ for the state (\ref{rij2a}). $D$ is maximum at
$\theta=\theta_m\approx 1.15\pi/4$. Bottom: The conditional entropies
$S(\rho'_{ij})-S(\rho'_{j})$ (upper curve) and $S(\rho_{ij})-S(\rho_j)$ (lower
curve) as a function of $\theta$, whose difference is the discord of the top
panel.}
 \label{f2} 
\end{figure}

In the general case, the eigenvalues of (\ref{rij2}) are $\lambda_{ij}=
(\half(1\pm \cos^2\theta),0,0)$, with those of $\rho_j(\theta)$ given by
$\lambda_j=\half(1\pm\cos\theta)$, whereas the ensuing eigenvalues (\ref{lmn})
of $\rho'_{ij}$ become, for $\cos 2\phi=1$,

\begin{figure}[t]

\centerline{\hspace*{0.25cm}\scalebox{.5}{\includegraphics{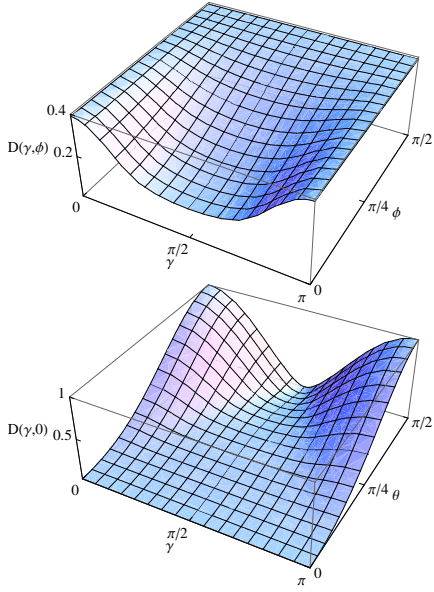}}}
\vspace*{-.25cm}

\caption{(Color online) Top: The difference (\ref{DS}) for the mixture
(\ref{rij2a}) as a function of $\gamma$ and $\phi$ at fixed $\theta=\pi/4$
(top) and as a function of $\gamma$ and $\theta$ at $\phi=0$ (bottom).
$D(\gamma,\phi)$ is minimum at $\gamma=\pi/2$ and $\phi=0$ $\forall$
$\theta\in(0,\pi/2]$.}
 \label{f1} 
\end{figure}

\begin{equation}
\lambda_{ij}'={\textstyle\frac{1+\nu\cos\theta\cos\gamma+\mu
\sqrt{[(1+\nu\cos\theta\cos\gamma)^2\cos^2\theta+
 \sin^2\gamma\sin^4\theta}}{4}}
 \label{lnu}\end{equation}
with those of $\rho'_j$ given by $\lambda_{j}'=
\half(1\pm\cos\theta\cos\gamma)$. For $\theta\in(0,\pi/2]$, the minimum of
(\ref{DS}) is always attained at $\gamma=\pi/2$ (and $\cos 2\phi=1$), i.e., for
a local measurement along the $x$ axis, as seen in Fig.\ \ref{f1}, with
$D(\gamma,0)$ becoming quite flat for small $\theta$. Both $\gamma=\pi/2$ and
$\gamma=0$ are stationary points of $D$, with $0$ a maximum. Fig.\ \ref{f2}
depicts the minimum $D\equiv D(\pi/2,0)$, which is the discord, as a function
of $\theta$, given explicitly by

\begin{eqnarray}
D&=&\sum_{\mu=\pm 1} \{[2h({\frac{1+\mu\sqrt{1-\frac{1}{4}\sin^2 2\theta}}{4}})
-h(\half)]\nonumber\\&&-[h({\frac{1+\mu\cos^2\theta}{2}})-
h({\frac{1+\mu\cos\theta}{2}})]\}\,,
 \label{Deth}\end{eqnarray}
where $h(x)=-x\log_2 x$. $D$ is maximum at $\theta=\theta_m \approx 1.15\pi/4$,
where $D\approx 0.15$. As seen in the bottom panel, while
$S(\rho'_{ij})-S(\rho'_j)$ (the first bracket in (\ref{Deth})) is an even
function of $\theta-\pi/4$, $S(\rho_{ij})-S(\rho_j)$ (the last bracket) is not,
being maximum at $\theta\approx 0.91\pi/4$ and originating the deviation of
$\theta_m$ from $\pi/4$.

For $\theta\rightarrow 0$, $D$ vanishes quadratically ($D\approx\half
\theta^2$) while for $\theta\rightarrow \frac{\pi}{2}$, $D\approx -\frac{1}{4}
(\frac{\pi}{2}-\theta)^2[\log_2(\frac{\pi}{2}-\theta)^2+\log_2 e-2]$.

\subsection{Effects of coherence term \label{CT}}
The reduced state of two spins at the factorizing field actually contains a
small coherence term $\propto \varepsilon(|\theta\theta\rangle\langle\!
-\theta\!-\!\theta|+|-\theta\!-\!\theta\rangle\langle\theta\theta|)$ (see sec.\ \ref{III}), which leads to the state
\begin{eqnarray}
\rho_{ij}^\varepsilon(\theta)&=&\frac{|\theta\theta\rangle\langle\theta\theta|+
|\!-\!\theta\!-\!\theta\rangle\langle\!-\!\theta\!-\!\theta|+
\varepsilon(|\theta\theta\rangle\langle\!-\!\theta\!-\!\theta|+h.c.)}
{2(1+\varepsilon\langle\theta\theta|\!-\!\theta\!-\!\theta\rangle)}
\label{rij3}\\
&=&\left(\begin{array}{cccc} a&0&0&\alpha\\0&\beta&\beta&0\\
0&\beta&\beta&0\\\alpha&0&0&b\end{array}\right)\,,\;
\begin{array}{l}(^a_b)=\frac{(1+\varepsilon)(1\pm\cos\theta)^2}
{4(1+\varepsilon\cos^2\theta)}\\
(^\alpha_\beta)=\frac{(1\pm\varepsilon)\sin^2\theta}
{4(1+\varepsilon\cos^2\theta)}\end{array}\,,
\end{eqnarray}
where $|\varepsilon|\leq 1$. This term generates then a parity dependent
correction to the previous results. The eigenvalues of $\rho_{ij}$ and $\rho_j$
are now $\lambda_{ij}= (a+b,2\beta,0,0)$, $\lambda_j=(a+\beta,b+\beta)$,
whereas those of $\rho'_{ij}$ and $\rho'_j$ can be obtained from Eq.\
(\ref{lmn}). The minimum of (\ref{DS}) is again obtained at $\gamma=\pi/2$ (and
$\phi=0$), (the surface being again similar to that of Fig.\ \ref{f1}), leading
to
\begin{eqnarray}
D&=&{\sum_{\mu=\pm 1}\{[
2h(\frac{1}{4}+\mu\frac{\sqrt{\cos^2\theta(1+\varepsilon)^2+\sin^4\theta}}
{4(1+\varepsilon\cos^2\theta)})
-h(\half)]}\nonumber\\&&{-[h(\frac{(1+\mu\cos^2\theta)(1+\mu\varepsilon)}
{2(1+\varepsilon\cos^2\theta)})
-h(\frac{(1+\mu\cos\theta)(1+\mu\varepsilon\cos\theta)}
{2(1+\varepsilon\cos^2\theta)})]\}\,.}\label{Deth2}
\end{eqnarray}
For $\varepsilon\neq 0$ a nonzero entanglement of the pair also arises, with
concurrence
\begin{equation}
C={\frac{|\varepsilon|\sin^2\theta}{1+\varepsilon\cos^2\theta}}\,,\label{C2}
\end{equation}
which is parallel (antiparallel) for $\varepsilon>0$ ($<0$).

In the limit $\varepsilon\rightarrow \pm 1$, Eq.\ (\ref{rij3}) becomes a pure
state, namely, $\rho_{ij}\rightarrow|\Psi_\pm\rangle\langle\Psi_\pm|$ with
\[|\Psi_\pm\rangle={\frac{|\theta\theta\rangle\pm
|-\theta-\theta\rangle}{\sqrt{2(1\pm\cos^2\theta)}}}=
\left\{\begin{array}{l}\frac{\cos^2\frac{\theta}{2}|00\rangle+
\sin^2\frac{\theta}{2}|11\rangle}{\sqrt{(1+\cos^2\theta)/2}}\\
 \frac{|01\rangle+|10\rangle}{\sqrt{2}}\end{array}\right.\,.\]
Therefore, $D$ and $E$ merge $\forall$ $\theta$ in this limit. Whereas
$|\Psi_-\rangle$ is a Bell state independent of $\theta$  (for $\theta\neq 0$),
leading to $D=E=C=1$, the entanglement of $|\Psi_+\rangle$ depends on $\theta$
(with $C=\sin^2\theta/(1+\cos^2\theta) \leq 1$), increasing with increasing
$\theta\in[0,\pi/2]$.

\begin{figure}[t]

\centerline{\hspace*{0.25cm}\scalebox{.45}{\includegraphics{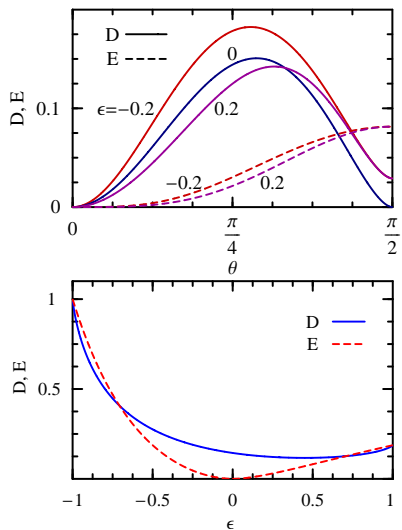}}}
\vspace*{-.25cm}

\caption{(Color online) Effect of coherence term. The quantum discord (D) and
entanglement of formation (E) of the state (\ref{rij3}) as a function of
$\theta$ for $\varepsilon=0.2$, $0$ and $-0.2$ (top) and as a function of
$\varepsilon$ at $\theta=\pi/4$ (bottom). Entanglement vanishes at
$\varepsilon=0$ but becomes larger than the discord close to the pure state
limit $\varepsilon=\pm 1$, where $D$ and $E$ coincide.}
 \label{f3} 
\end{figure}

The response of $D$ and $E$ to the coherence term is shown in Fig.\ \ref{f3}.
For sufficiently small $\varepsilon$, the correction to $D$ is linear in
$\varepsilon$ for $\theta$ not close to $\pi/2$, with the discord increasing
(decreasing) for $\varepsilon<0$ ($>0$), while at $\theta=\pi/2$ the correction
is quadratic and positive (at $\theta=\pi/2$, $D=1-\sum_{\mu=\pm
1}h(\frac{1+\mu\varepsilon}{2})\approx\half\varepsilon^2\log_2 e$).
Entanglement, on the other hand, becomes finite as soon as $|\varepsilon|$
increases, becoming even larger than the discord for $\theta$ close to $\pi/2$
(where $C=|\varepsilon|$ and $E\propto-\varepsilon^2\log\varepsilon^2$ for
small $\varepsilon$). As seen in the bottom panel, at an intermediate $\theta$
entanglement remains smaller than the discord just in an interval around
$\varepsilon=0$, becoming slightly larger before reaching the pure limit
$\varepsilon=\pm 1$, where $D$ and $E$ coincide.

\section{Quantum Discord in XY chains\label{III}}
We have now all the elements to evaluate and understand the behavior of the
discord of an arbitrary spin pair in the ground state of a finite chain. We
first consider a finite $1$-$d$ $XY$ cyclic chain of $n$ spins with first
neighbor couplings, where $J_z^{ij}=0$ and
\begin{equation}
J_\mu^{ij}=\delta_{j,i\pm 1}J_\mu,\;\;\mu=x,y\,,\label{nn}
 \end{equation}
with $n+1\equiv 1$. The exact solution for finite $n$ can be obtained through
the Jordan-Wigner fermionization \cite{LM.61} (see Appendix). The exact ground
state will have a definite (field dependent) spin parity and the reduced
density of an arbitrary pair will  be of the form (\ref{rij}), where the
elements (\ref{vm1})-- (\ref{vm3}) can be evaluated with the expressions of the
Appendix.

\begin{figure}[t]
\vspace*{0cm}

\centerline{\hspace*{0.cm}\scalebox{.7}{\includegraphics{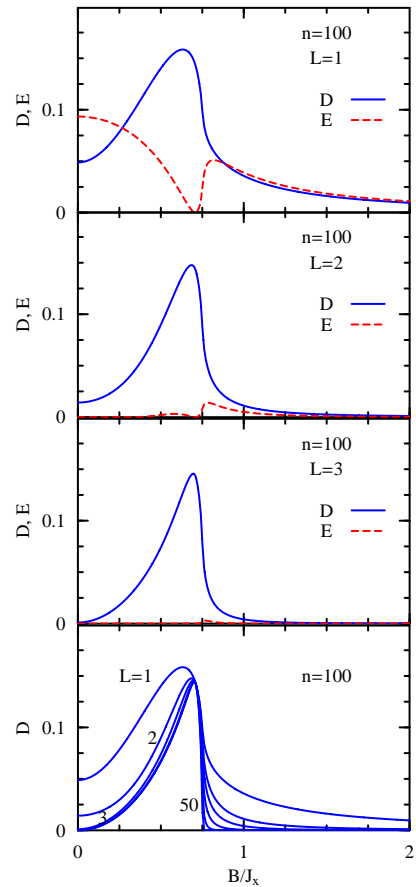}}}
 \vspace*{-0.5cm}

\caption{(Color online) Quantum discord (D) and entanglement (E) of spin pairs
with separation $L$ in the exact ground state of a cyclic chain of $n=100$
spins with first-neighbor $XY$ couplings and $J_y/J_x=0.5$, as a function of
the transverse magnetic field. The top panel corresponds to first neighbors
($L=1$). At the factorizing field $B_s=\sqrt{J_yJ_x}\approx 0.71 J_x$, $E$
vanishes whereas $D$ approaches the same finite limit (\ref{Deth}) $\forall$
$L$, with $\theta$ determined by (\ref{cth}). Bottom: Discord for all
separations $L$. A finite saturation limit is approached for large $L$ if
$|B|<B_c$.}
 \label{f4}
\end{figure}

\begin{figure}[t]
\vspace*{0.cm}

\centerline{\hspace*{0.cm}\scalebox{.7}{\includegraphics{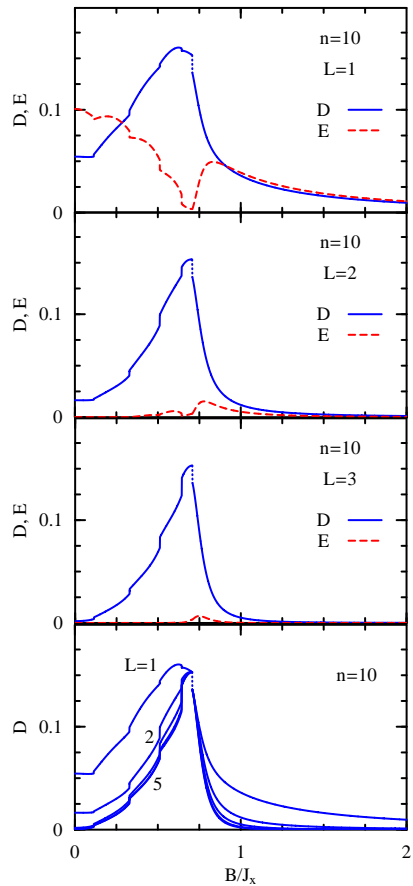}}}
 \vspace*{-0.5cm}

\caption{(Color online) Same details as Fig.\ \ref{f4} for a chain with $n=10$
spins. The common different side-limits of $D$ at the factorizing field, given
by Eq.\ (\ref{Deth2}) with coherence factor (\ref{eps}), are now appreciable.}
 \label{f5}
\end{figure}

Exact results for the discord and entanglement of pairs are shown in Figs.\
\ref{f4}--\ref{f5} for $\chi=J_y/J_x=0.5$ and two different sizes $n$. For a
first neighbor coupling and even $n$, the sign of $J_x$ can be changed by a
local transformation $s_{2i,\mu}\rightarrow -s_{2i,\mu}$ for $\mu=x,y$, so that
both the ferromagnetic ($J_x>0$) and antiferromagnetic ($J_x<0$) cases at fixed
$\chi$ will exhibit exactly the same entanglement and discord. They are also
independent of the sign of $B$.

It is immediately seen that pair entanglement and discord exhibit significant
differences for fields $|B|<B_c=\half(1+\chi)J_x$ (the critical field of the
thermodynamic limit $n\rightarrow\infty$). While in this case entanglement
practically vanishes at the factorizing field
\cite{Kur.82,Ros.04,Am.06,RCM.08,GAI.09}
 \begin{equation}B_s=\sqrt{J_y J_x}\,,\label{Bs}\end{equation}
where the chain possesses an exactly separable and degenerate parity-breaking
ground state \cite{RCM.08} (see Eq.\ (\ref{mfth})), the discord remains
non-zero, reaching in fact its maximum in its vicinity. Besides, entanglement
rapidly decreases as the separation $L$ between the spins increases, being
nonzero for $L>2$ only in the immediate vicinity of $B_s$, where it is very
small. In contrast, the discord decreases only slightly with separation for
$|B|<B_c$, reaching a saturation value for large $L$. Moreover, it is strictly
{\it independent} of $L$ at the factorizing field $B_s$.

In order to understand these results, we recall that at $B=B_s$, the uniform
parity-breaking separable state
\begin{eqnarray}
|\Theta\rangle&=&\otimes_{i=1}^n|\theta_i\rangle,\;\;|\theta_i\rangle=
\exp[i\theta s_{iy}|0_i\rangle\,,\label{mfth}\\
\cos\theta&=&B_s/J_x=\sqrt{\chi}\,,\label{cth}
 \end{eqnarray}
where $s_{iz}|0_i\rangle=-\half |0_i\rangle$, is an {\it exact} ground state 
(we have here assumed $J_x>0$; for $J_x<0$, $\theta_i\rightarrow (-1)^i \theta$). It is obviously a state with spins fully aligned along an axis forming an angle $\theta$ with the original $-z$ axis. Due to parity symmetry, the partner state $|-\Theta\rangle=P_z|\Theta\rangle$ is also an exact ground state at $B_s$. They can therefore be exact eigenstates of $H$ only when levels of opposite parity cross \cite{RCM.08}. The ground state of the present chain undergoes actually $n/2$ spin parity transitions as the field $B$ increases from $0$ (reminiscent of the $m/2$ $S_z$ transitions of the $XX$ limit \cite{CR.07}), the last one precisely at $B_s$. Outside these transitions, the energy gap between the lowest states of each parity is small for $|B|\alt B_c$, but nonetheless {\it finite} in a finite chain.

In the immediate vicinity of $B_s$, the exact ground state will therefore
possess a definite parity and the correct side-limits at $B_s$ will be
determined by the states
\begin{equation}
|\Theta^{\pm}\rangle=\frac{|\Theta\rangle\pm|\!\!-\Theta\rangle}{
\sqrt{2(1\pm\langle -\Theta|\Theta\rangle)}}\,, \label{thp}
\end{equation}
where $\langle\!-\Theta|\Theta\rangle=\cos^n\theta$ \cite{RCM.08}. The reduced
state of {\it any} two spins derived from (\ref{thp}) will be given precisely
by the same mixture (\ref{rij2}) if the complementary overlap $\langle -\Theta_{n-2}
|\Theta_{n-2}\rangle=\cos^{n-2}\theta$ is neglected. We can then immediately
understand the qualitative difference between entanglement and discord for
$|B|<B_c$. As $B$ approaches $B_s$, the discord between {\it any} two spins will
approach {\it a common finite limit for {\it any} separation $L$}, given by
Eq.\ (\ref{Deth}) with $\theta$ obtained from (\ref{cth}) ($D\approx 0.145$ in the case of
fig.\ \ref{f4}, where $\theta=\pi/4$ at $B_s$). In contrast, the pair
entanglement will vanish for $B\rightarrow B_s$ (for a negligible complementary 
overlap) as the state (\ref{rij2a}) is separable. 

\begin{figure}[t]
\vspace*{0.cm}

\centerline{\hspace*{0.cm}\scalebox{.5}{\includegraphics{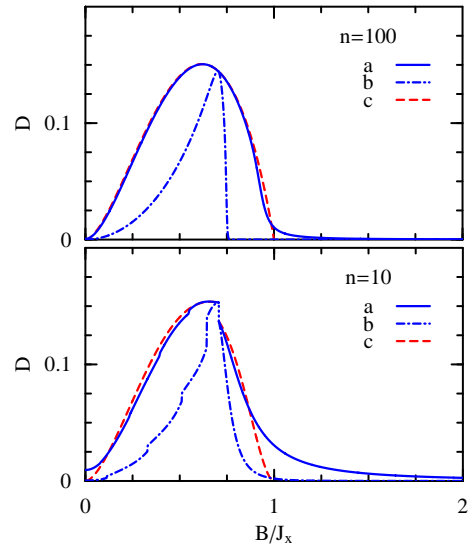}}}
 \vspace*{-0.5cm}

\caption{(Color online) Top: Discord between spin pairs in the $n=100$ fully
connected array (a), together with the discord (\ref{Deth}) of the state
(\ref{rij2a}) at the mean field angle $\cos\theta=B/J_x$ (c). The result for
first neighbor coupling at the same anisotropy and size $n$ is also depicted
(b), for large separation ($L=n/2$). The limits at the factorizing field $B_s$
are exactly coincident. Bottom: Same details for $n=10$ spins. Here $(c)$
depicts the result (\ref{Deth2}) for the actual mixture (\ref{rij3}), which
includes the coherence term. Side limits at $B_s$ are again coincident.}
 \label{f6}\vspace*{-.2cm}
\end{figure}

In Figs.\ \ref{f4}-\ref{f5} we have taken the exact ground state with its
correct parity. The non-negligible discord between any two spins for $|B|<B_c$
can then be understood in a similar way, as in this region the ground state can
be seen, approximately, as a definite parity combination (\ref{thp}) of ``mean
field'' states with broken-symmetry (\ref{mfth}), with $\cos\theta=B/J_x$, plus
additional corrections. The reduced state of a spin pair will then be again
given essentially by the mixture $\rho(\theta)$ (Eq.\ (\ref{rij2a})) plus
smaller corrections, with the discord arising principally from $\rho(\theta)$
(although corrections are non-negligible; see Fig.\ \ref{f6}).

Let us remark that the same reduced density (\ref{rij2a}) arises from the
statistical mixture $\rho_0=\half(|\Theta^+\rangle\langle\Theta^+|
+|\Theta^-\rangle\langle\Theta^-|)$ if the overlap is discarded. $\rho_0$
represents the $T\rightarrow 0^+$ limit of the thermal state of the system at 
$B_s$. The limit (\ref{Deth}) of the discord at $B_s$ remains then also valid
at sufficiently low $T$. 

On the other hand, for strong fields $B\gg J_x$, the ground state is
essentially the state with all spins fully aligned along the $-z$ axis plus
small perturbative corrections. As seen in Fig.\ \ref{f4}, the discord in this
region is rather small and decreases rapidly with separation, since the
previous superposition effects are no longer present. Moreover, the discord
between first neighbors is very close to the entanglement of formation, as
verified by a perturbative expansion: For $|B|\gg J_x$ and $L=1$ in the case
(\ref{nn}) we obtain, setting $\eta=(J_x-J_y)/(8B)$,
\begin{eqnarray}
D&\approx& {\textstyle \eta^2(-\log_2\eta^2+\log_2 e-2)}\\
E&\approx & {\textstyle \eta^2(-\log\eta^2+\log_2 e)}\,.
\end{eqnarray}
Thus, $E$ is in this region slightly {\it greater} than $D$, as verified in the
top panel of Figs.\ \ref{f4}--\ref{f5}.

Results for a small chain of $n=10$ spins are shown in Fig.\ \ref{f5}. Although
the behavior is similar to that for $n=100$, finite size effects become
important and the overlap $\langle-\Theta |\Theta\rangle$ can no longer be
neglected. The effects on the discord and entanglement of the ground state
parity transitions taking place for $|B|<B_s$ are now visible, giving rise to
small steps in these quantities. The final step takes place at $B_s$, where $D$
exhibits now a {\it finite discontinuity} due to the parity splitting arising
from the coherence term (sec.\ \ref{CT}), no longer negligible: the actual
reduced state of a spin pair derived from the states (\ref{thp}) is given by
Eq.\ (\ref{rij3}) with
\begin{equation}
\varepsilon=\pm\cos^{n-2}\theta\,,\label{eps}
\end{equation}
where the $-$ ($+$) sign corresponds to the left (right) side at $B_s$, i.e.,
negative (positive) spin parity. The side-limits of $D$ at $B_s$ are then given
by Eq.\ (\ref{Deth2}) for the values (\ref{cth})--(\ref{eps}) of $\theta$ and
$\varepsilon$ (leading to $D_-\approx 0.153$, $D_+\approx 0.137$ in the case of
Fig.\ \ref{f5}). Small but non-zero common side-limits at $B_s$ of the
entanglement between any two spins also arise \cite{RCM.08}, as determined by
Eqs.\ (\ref{Eij}), (\ref{C2}). In contrast with the discord, Eq.\ (\ref{C2}) satisfies of course the monogamy inequality \cite{CKW.00,OV.06}, reaching its maximum value $2/n$ in the $XX$ limit $J_y\rightarrow J_x$ (where $\theta\rightarrow 0$ and $C(\varepsilon)\rightarrow 0$ but $C(-\varepsilon)\rightarrow 2/n$, 
as $|\Theta^-\rangle$ approaches the $W$-state \cite{RCM.08}). 

The present values of the discord at the factorizing field, determined by Eqs.\
(\ref{Deth}) or in general (\ref{Deth2}), are actually much more general:
Uniform chains or arrays with ferromagnetic ($J_x^{ij}>0$) $XY$ couplings of {\it arbitrary} range but {\it common} anisotropy $\chi=J_y^{ij}/J_x^{ij} \in(0,1)$ will also exhibit a factorizing field, given for spin $1/2$ by  \cite{RCM.08,GAI.09}
 \begin{equation} B_s=\sqrt{\chi}\sum_{j\neq i} J^{ij}_x\,,\label{Bs2}\end{equation}
where the chain will possess again the same degenerate separable ground states
(\ref{mfth}). Side-limits at $B_s$  will then be determined by the same
definite parity states (\ref{thp}). The same occurs in $XYZ$ arrays if
$\chi=(J^{ij}_y-J^{ij}_z)/(J^{ij}_x-J^{ij}_z)$ is constant \cite{RCM.08}, with 
$B_s=\sqrt{\chi}\sum_{j\neq i}(J^{ij}_x-J_z^{ij})$. As a result, the ground state pair discord in all 
these systems will be finite and independent of pair separation or coupling
range in the vicinity of $B_s$, and given by Eqs.\ (\ref{Deth}) or
(\ref{Deth2}), with the values (\ref{cth})--(\ref{eps}). Similar arguments will
apply in the vicinity of more general factorizing fields \cite{RCM.09}.

An example is provided in Fig.\ \ref{f6}, where results for a fully and
uniformly connected $XY$ array (LMG model \cite{LMG.65}),
 \begin{equation}
 J_\mu^{ij}=(1-\delta_{ij})J_\mu/(n-1)\,,\;\;\mu=x,y\,,
 \end{equation}
with the same anisotropy $J_y/J_x=0.5$, are depicted. The exact result can be
here obtained by direct diagonalization, as $H$ can be expressed in terms of
the total spin operators $S_\mu=\sum_i s_{i\mu}$. The reduced pair density  will obviously be independent of separation at any field. The pairwise
entanglement is then small for large $n$, with a $O(n^{-1})$ concurrence
\cite{RCM.08}.

The discord is, however, non-negligible and practically $n$-independent for
large $n$. It is verified in Fig.\ \ref{f6} that at $B=B_s$, the same previous
limit (\ref{Deth}) ($n=100$) and side-limits (\ref{Deth2})--(\ref{eps})
($n=10$) at $B_s$ are obtained. Moreover, the simple model (\ref{rij2a}) (large
$n$) or (\ref{rij3}) (small $n$) for the reduced density of a spin pair
accurately describes the discord {\it in the whole region $|B|<B_c=J_x$} (and not just at $B_s$) if the mean field value $\cos\theta=B/J_x$ is employed for
$\theta$, as seen in Fig.\ \ref{f6}. This indicates that the effects on $D$ of
correlations beyond the basic mean field description with parity restoration of
the ground state given by Eq.\ (\ref{thp}), become in this system very small, being negligible for large $n$. In contrast, such mean field model cannot accurately describe the pair discord (either for small or large separations) in the nearest-neighbor chain away from $B_s$.

\section{Conclusions \label{IV}}
We have first discussed in detail the discord of a mixture of aligned pairs in
two different directions. While exhibiting no entanglement if coherence effects
between both directions are negligible, these mixtures do exhibit a finite and
non-negligible quantum discord if the directions are not coincident nor
opposite. In the presence of coherence terms, however, entanglement becomes
finite and can be larger than the discord.

Such mixtures become of crucial importance for studying the pair discord in the
exact (and hence of definite parity) ground state of finite $XY$ and $XYZ$ spin
chains in a transverse field. They represent the actual reduced state of an
arbitrary pair in the vicinity of the factorizing field $B_s$. Previous results
imply then a finite discord between {\it any} two spins in the vicinity of
$B_s$, {\it irrespective of pair separation or coupling range}. These mixtures
are also the main term of the reduced pair state in the whole region $|B|<B_c$,
implying there a non-negligible pair discord even for pairs not linked by the
couplings, as was seen in the nearest-neighbor case. Such mixtures do in fact
accurately describe the pair discord $\forall$ $|B|<B_c$ in the fully connected
$XY$ model. 

The behavior of the discord, which is free from the monogamy restriction, 
differs then considerably from that of the pairwise entanglement, whose limits at $B_s$ are 
small and determined by the coherence term. This term gives rise to a parity splitting and hence to a finite discontinuity of the discord at $B_s$, visible in small chains.

A final remark is that the present results, together with those previously
obtained for the entanglement \cite{RCM.08}, allow to identify the factorizing
field as a quantum critical point for the ground state of an $XY$ or $XYZ$  chain of small size: At $B_s$, the last ground state parity transition takes place and in its immediate vicinity, pair quantum correlations, as measured by the discord, become independent of both pair separation and coupling range.

The authors acknowledge support of CONICET (LC, NC) and CIC (RR) of Argentina.

\appendix
\section{Exact solution of the finite cyclic XY chain}
The Jordan-Wigner transformation \cite{LM.61} allows to rewrite Eq.\ (\ref{H})
for the case of first neighbor couplings (\ref{nn}) and  {\it for each value
($\pm$) of the spin parity $P_z$} (Eq.\ (\ref{Pz})), as a quadratic form in
fermion operators $c^\dagger_i$, $c_i$ defined by $c^\dagger_i=s_i^+\exp[-i\pi
\sum_{j=1}^{i-1}s_j^+s_j^-]$:
\begin{eqnarray} H^{\pm}&=&
\sum_{i=1}^n B(c^\dagger_ic_i-\half)-\half\eta^{\pm}_i(J_+c^\dagger_i c_{i+1}
+J_-c^\dagger_i c^\dagger_{i+1}+h.c.)\nonumber\\
&=&\sum_{k\in K_{\pm}}\!\!\lambda_k (a^\dagger_k a_k
 -\half),\;\label{qd}\end{eqnarray}
where $J_{\pm}=\half(J_x\pm J_y)$ and, in the cyclic case $n+1\equiv 1$,
$\eta^-_i=1$, $\eta^+_i=1-2\delta_{in}$ \cite{LM.61}. In (\ref{qd}),
 \[\lambda_k^2=(B-J_+\cos\omega_k)^2+J_-^2\sin^2\omega_k, \;\;\;
 \omega_k=2\pi k/n\,, \]
with $K_+=\{\half,\ldots,n-\half\}$, $K_-=\{0,\ldots,n-1\}$, i.e., $k$ {\it
half-integer (integer) for positive (negative) parity} \cite{RCM.08,CR.07}. The
diagonal form (\ref{qd}) is obtained through a discrete {\it parity-dependent}
Fourier transform $c^{\dagger}_j=\frac{e^{i\pi/4}}{\sqrt{n}} \sum_{k\in
K_{\pm}} e^{-i\omega_k j}c'^\dagger_k$, followed by a BCS transformation
$c'^\dagger_k=u_k a^\dagger_k+v_ka_{n-k}$, $c'_{n-k}=u_k a_{n-k}-v_k
a^\dagger_k$ to quasiparticle fermion operators $a_k$, $a^\dagger_k$, with
$u_k^2,v_k^2=\half[1\pm(B-J_+\cos\omega_k)/\lambda_k]$. For $B\geq 0$ we set
$\lambda_k\geq 0$ for $k\neq 0$ and $\lambda_0=J_+-B$, such that the
quasiparticle vacuum in $H^-$ is odd and the lowest energies for each parity
are $E^{\pm}=-\half\sum_{k\in k_{\pm}}\lambda_k$. At $B=B_s=\sqrt{J_xJ_y}$,
$\lambda_k=J_+-B_s\cos\omega_k$ and $E^\pm=-nJ_+/2$ \cite{RCM.08}.

The concurrences in the fixed parity GS can be obtained from the contractions
$f_l\equiv\langle c^\dagger_ic_j\rangle_\pm-\half \delta_{ij}$,
$g_l\equiv\langle c^\dagger_i c^\dagger_j\rangle_\pm$ and the use of Wick's
theorem \cite{LM.61}, leading to $\langle s_z^i\rangle=f_0$, $\langle
s_z^is_z^j\rangle=f_0^2-f_l^2+g_l^2$ and
$\alpha_l^{\pm}={\textstyle\frac{1}{4}[{\rm det}(A^+_l) \mp{\rm det}(A^-_l)]}$,
with $(A_l^\pm)_{ij}=2(f_{i-j\pm 1}+g_{i-j\pm 1})$ $l\times l$ matrices. All
previous results have been explicitly checked for small $n$ with those obtained
from a direct diagonalization.

\end{document}